\documentclass[11pt,a4paper]{article}
\usepackage[hyperref]{acl2019}
\usepackage{times}
\usepackage{latexsym}
\usepackage{amsmath}
\usepackage{url}

\aclfinalcopy 

\title{Discourse Behavior of Older Adults Interacting With a Dialogue Agent Competent in Multiple Topics}

 \author{S. Zahra Razavi \\
  University of Rochester \\
  Rochester, NY, USA \\
  \texttt{srazavi@cs.rochester.edu} \\\And
  Lenhart K. Schubert \\
  University of Rochester \\
  Rochester, NY, USA \\
  \texttt{schubert@cs.rochester.edu} \\
  \AND
  Kimberly A. Van Orden \\
  University of Rochester Medical Center \\
  Rochester, NY, USA \\
  \texttt{kimberly\_vanorden@urmc.rochester.edu} \\\And
  Mohammad Rafayet Ali \\
  University of Rochester \\
  Rochester, NY, USA \\
  \texttt{\hspace{.9cm}mali7@cs.rochester.edu} \\
  }

\date{}

\begin{document}
\maketitle
\begin{abstract}
  
  We present some results concerning the dialogue behavior and inferred sentiment of a group of older adults interacting with a computer-based avatar. Our avatar is unique in its ability to hold natural dialogues on a wide range of everyday topics---27 topics in three groups, developed with the help of gerontologists. The three groups vary in ``degrees of intimacy", and as such in degrees of difficulty for the user. Each participant interacted with the avatar for 7-9 sessions over a period of 3-4 weeks; analysis of the dialogues reveals correlations such as greater verbosity for more difficult topics, increasing verbosity with successive sessions, especially for more difficult topics, stronger sentiment on topics concerned with life goals rather than routine activities, and stronger self-disclosure for more intimate topics. In addition to their intrinsic interest, these results also reflect positively on the sophistication of our dialogue system.
  
\end{abstract}

\section{Introduction}

\vspace*{-.06in}

Spoken dialogue systems have proved beneficial for helping older people with their needs, including social companionship \cite{miehle2019social,abdollahi2017pilot}, health advice \cite{onovirtual}, palliative care \cite{utami2017talk}, reminiscence therapy \cite{arean1993comparative} and many other applications. In many of these applications, it is crucial to keep users involved in the task for multiple sessions through engaging conversations. 
Here we seek insights into user behavior when interacting with a system capable of conversing about various casual topics.

Our study is based on linguistic 
data collected from 80 sessions of interaction with nine participants over a course of 3-4 weeks. The topics were suggested by gerontological experts and are categorized into three groups based on their degree of intimacy. We investigate how users' verbosity, sentiment, and self-disclosure behavior depend on the topic under discussion and the avatar's tone, and how they evolve with time. We measure verbosity in terms of users' turn lengths, sentiment in terms of the Vader sentiment analysis tool, and self-disclosure in terms of cues suggested by the literature, and extracted using LIWC categories. We present the insights obtained from the analysis and discuss the results.

\section{Background}
\label{sec:background}

\vspace*{-.06in}

Content analysis of dialogues with non-task oriented conversational agents (CAs) has proved helpful in increasing CAs' effectiveness in a variety of tasks. For instance, detecting the main themes of a dialogue, or speech and language features, could assist in detecting schizophrenia \cite{dellazizzo2018exploration} and dementia \cite{ujiro2018detection}, and preventing suicide \cite{martinez2017embodied}. 
An important aspect of conversational content we investigate in this paper is the degree of self-disclosure. Encouraging self-disclosure can increase rapport in user-CA interaction \cite{pecune2018field} and thereby the effectiveness of the virtual agent in different tasks (e.g., health coaching \cite{lisetti2013can}). We also study the role of sentiment, since use of sentiment features has been observed to increase the quality of conversational agent output \cite{rinaldi2017end}.
While sentiment can be evaluated using analysis tools such as Vader \cite{hutto2014vader}, the set of features indicative of self-disclosure remains ill-defined. \cite{ravichander2018empirical} suggested utterance length, negation words, POS tags, and emotion-laden words as self-disclosure markers in an open-ended conversation with a chatbot. \cite{houghton2012linguistic} identifies personal pronouns, word count, and family and sexual words as significant, based on comparing secret tweets with normal tweets. \cite{bak2014self} observed that tweets with deeper self-disclosure contain secretive wishes or sensitive information while medium self-disclosing tweets convey general information about self such as family, education, etc. In our work we tried several LIWC categories based on the cited literature.

\vspace*{-.06in}

\section{Dataset}
\label{sect:dataset}

\vspace*{-.06in}

Our data came from multi-session interactions between a screen-based virtual agent and elderly users, where the agent leads users in casual conversations controlled by an automatic dialogue manager. The system was designed as a tool allowing users to practice their communication skills, giving them feedback on their non-verbal behavior and speech prosody. We recruited nine participants, each of whom had seven to nine sessions with the avatar; the first and the last interaction were held in the lab and the rest were self-initiated by users at home. Participants were asked to fill out surveys and were evaluated for their communication skills by experts.\\*[.08in] 
%
{\bf Dialogues.} Each interaction consists of 3 subsessions, each containing 3-5 questions from the avatar on a specific topic listed in table \ref{topic-table}. The dialogue manager follows a plan for each topic, asks some questions, extracts essential information from users' inputs and produces relevant comments indicating its understanding of the user. Each interaction took 15-20 minutes depending on the number of questions and the user's verbosity. 

The topics were selected by gerontological experts and divided into three groups based on their emotional intensity: easy, medium, and hard. ``Easy" (less intimate) topics are ones likely to be broached in making someone's acquaintance, while the harder ones are more emotionally evocative and call for more self-disclosure. As can be seen in table \ref{topic-table}, the dialogue sessions were designed so that users start with easier topics in earlier sessions and gradually transition to harder ones as they progress in the study.

\begin{table}[t!]
\begin{center}
\scalebox{0.9}{
\begin{tabular}{|l|ll|}
\hline \textbf{} & \textbf{Subsessions Topics} & \textbf{EI} \\ \hline
S1 & Getting to know (I, II), Activity & E,E,E \\
S2 & City you live in (I, II), Pets & E,E,M \\
S3 & Family, Gathering, Yourself & E,M,H \\
S4 & Weather, Driving, Cooking & E,H,E \\
S5 & Outdoors, Travel, Plan for today & M,M,E \\
S6 & Chores, Money, Growing older & E,M,H \\
S7 & Education, Job, Life goals  & M,M,H\\
S8 & Technology, Books, Arts & M,M,M\\
S9 & Sleep, Health, Exercise & M,M,M\\

\hline
\end{tabular}}
\end{center}
\caption{\label{topic-table} Dialogue topics and their emotional intensity level (E: easy, M: medium, H: hard)}
\end{table}

\vspace*{.08in}
\noindent
{\bf Data statistics.} We collected the transcripts (produced via ASR) from nine users interacting with the system over seven to nine days.
A few subsessions were missed because of technical issues. Table ~\ref{data-stat-table} summarizes the collected data.

\begin{table}[t!]
\begin{center}
\scalebox{0.9}{
\begin{tabular}{|l|r|}
\hline \textbf{Feature} & \textbf{Number} \\ 
\hline
users interacted with the avatar & 9 \\
subsessions & 198 \\
total users' turns & 668 \\
total users' words & 29054 \\
total avatars' words & 24296 \\
\hline
\end{tabular}}
\end{center}
\caption{\label{data-stat-table} Collected data statistics }
\end{table}

\label{sect:pdf}

\section{Dialogue Content Analysis}

\vspace*{-.06in}

We analyzed three aspects of the dialogue content. The first concerns verbosity, where we looked for differences in verbosity across different sessions, users, and topic classes; we also analyzed changes in verbosity over time.
The second concerns the results of sentiment analysis for different sessions and the tone change over time. The final aspect concerns the kinds of self-disclosure cues we gleaned from the literature.

\subsection{Verbosity}

\begin{figure}
  \includegraphics[width=\columnwidth]{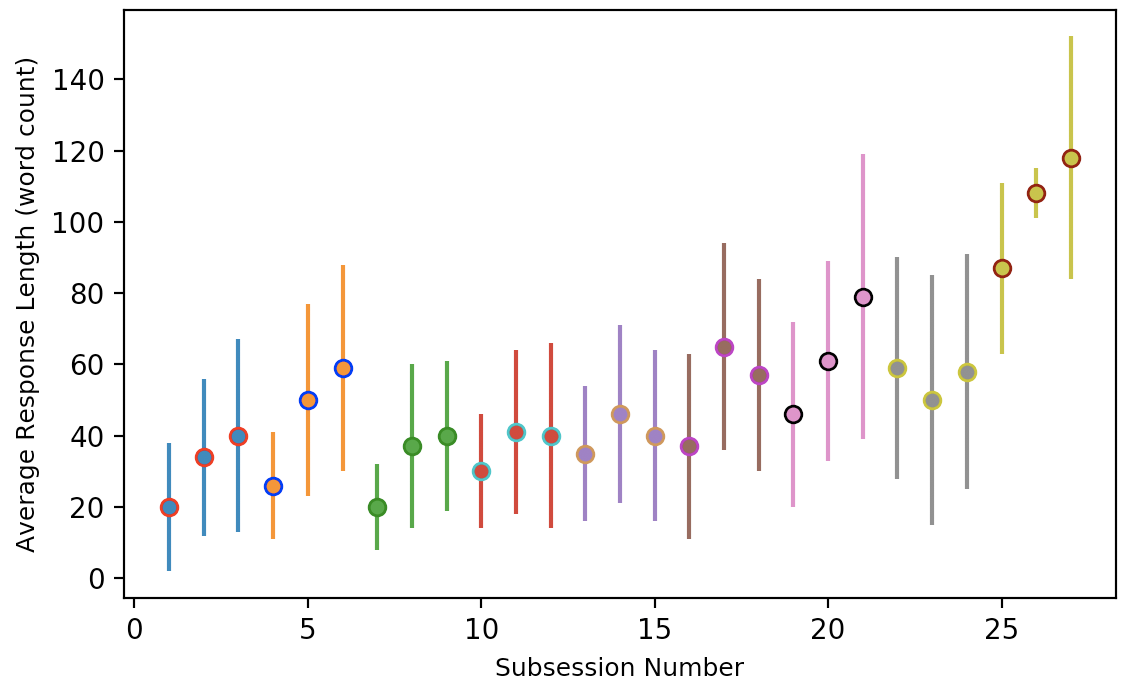}
  \caption{Users' average turn length in each subsession}
  \label{fig:wc-to-subsession}
\end{figure}

Our metric for utterance length was word count.\\*[.08in] 
\noindent
{\bf Response length change over time.}
The results show that users on average tend to provide longer responses as they proceed in a conversation. Figure \ref{fig:wc-to-subsession} shows average response length among all users in different subsessions. We also observe a strong, significant correlation (a) between the average word count and the particular subsession (Pearson $r=0.76$, $p<10^{-5}$); (b) between the average word count and the user's turn number in the whole interaction ($r=0.68$, $p<10^{-12}$; and (c) between the average word count and the interaction number ($r=0.81$, $p=0.008$). Trends (b) and (c), however, are not the same for all individuals. For five out of nine users the turn length correlation with time is significantly strong, while for the rest we cannot see any significant correlation.

\vspace*{.08in}
\noindent
{\bf Users' turn length and topic classes.}
The introduced topic classes significantly affect users' the response length. The average among all users shows that users provide longer responses to ``hard" questions, where the average is $57.60 (\sigma = 33.92)$ words, while responses to ``medium" and ``easy" questions contain an average number of $51.41 (\sigma = 30.39)$ words and $32.73 (\sigma = 24.03)$ words respectively.  Interestingly, the response length change over time is not significant for easy topics but it is significantly strong for medium ($r=0.81$, $p<10^{-3}$) and hard ($r=0.94$, $p=0.05$) topics.\\*[.08in]
\noindent
{\bf User and avatar turn length.}
Some studies suggest that the utterance lengths of one speaker can influence the interlocutor's utterance lengths. We looked for any correlation between the avatar's input length and users' corresponding turn length, but did not observe any meaningful relation.  
   
\subsection{Sentiment}
\begin{table}[t!]
\begin{center}
\scalebox{0.9}{
\begin{tabular}{|l|l|l|}
\hline  \textbf{Topic class} & \textbf{$ave(\lvert Sent_{a}\rvert)$} & \textbf{$ave(\lvert Sent_{u}\rvert)$} \\ \hline
Easy & 0.3 ($\sigma=0.31$) &  0.43 ($\sigma=0.35$) \\
Medium & 0.36 ($\sigma=0.3$) &  0.62 ($\sigma=0.31$) \\
Hard &  0.38 ($\sigma=0.32$) &  0.63 ($\sigma=0.3$) \\
\hline
\end{tabular}}
\end{center}

\caption{\label{topic-sent-table} Average sentiment score for topic classes}
\end{table}
We used VADER \cite{hutto2014vader} to quantify utterance sentiment for each avatar and user turn.\\*[.08in]
\noindent
{\bf User vs.\ avatar turn sentiment}
The correlation coefficient value shows a weak but significant correlation between a given user turn and the avatar's preceding turn ($r=0.23$, $p<10^{-8}$); this suggests a slight dependence of the user's sentiment on the avatar's tone (though both might be derivative from the particular question content).
In order to compensate for the possible influence of the avatar's tone on the user, we study sentiment difference over time ($Sentiment_{user} - Sentiment_{avatar}$). We observe a significant weak increase in positive tone over time ($r=0.35$, $p<10^{-3}$). \\*[.08in]
\noindent
{\bf Sentiment for different topics.}
A more careful look into different interaction sessions provides some insight into the relation between user sentiment and dialogue topics. We should first note that the avatar is designed to convey a positive, friendly tone in its interactions, thereby encouraging a generally positive tone on user's side. However, we find user sentiment to be significantly more positive for some topics than others. Among them are ``Travel", with sentiment score= $0.74 (\sigma=0.22)$, ``Health", with sentiment score= $0.76 (\sigma=0.13)$, ``Education", with sentiment score= $0.74(\sigma=0.07)$, and ``Outdoor", with sentiment score= $0.68(\sigma=0.28)$. On the other hand, in talking about subjects such as ``Family", ``Getting to know each other", and ``Managing money" people tend to be more neutral, with respective average sentiment scores of $0.045(\sigma=0.18)$, $0.19(\sigma=0.38)$, $0.32(\sigma=0.39)$. 

We infer that topics concerned with life goals evoke stronger emotions than those concerned with routine activities of daily life. As well, discussion of eventualities such as
the death of a partner or living alone after others have moved out naturally leads to a more negative emotional tone. There are other themes that evoke both negative and positive user comments, and hence sentiment fluctuations resulting in a high standard deviation and no meaningful average. An example is the topic ``Growing older" with sentiment score = $0.37 (\sigma=0.50)$.\\*[.08in]
\noindent
{\bf Sentiment for different topic classes.}
We also studied the average sentiment for the three topic classes introduced in section \ref{sect:dataset}. Our hypothesis was that emotionally evocative topics produce stronger user sentiment than more neutral ones. We therefore evaluated the average absolute sentiment value across all users for different topic classes. The results can be seen in table \ref{topic-sent-table}.

The results show that although the avatar's tone remains almost the same for all classes, users tend to use stronger tones when they talk about `medium' and `hard' topics compared to `easy' ones.   
\begin{table}[t!]
\begin{center}
\scalebox{0.9}{
\begin{tabular}{|l|l|l|l|}
\hline  \textbf{Potential SD Cues} & \textbf{Easy} & \textbf{Medium} & \textbf{Hard} \\ \hline
Word count per turn & 31.97 & 49.61 & 55.45 \\
1st person pron. & 9.91 & 9.29 & 9.46 \\
Family and friend & 1.02 & 1.08 & 1.03 \\
Negative emotion & 0.51 & 0.54 & 0.84 \\
Positive emotion &  4.91 & 4.9 & 5.54 \\
Drives & 5.71 & 6.82 & 7.26 \\
Personal concerns & 5.2 & 7.07 & 4.15 \\
\hline

\end{tabular}}
\end{center}
\caption{\label{sd-cues-score-table} LIWC score of SD cues for three topic classes}
\end{table}

\subsection{Self-disclosure}

\begin{figure}
  \includegraphics[width=\columnwidth]{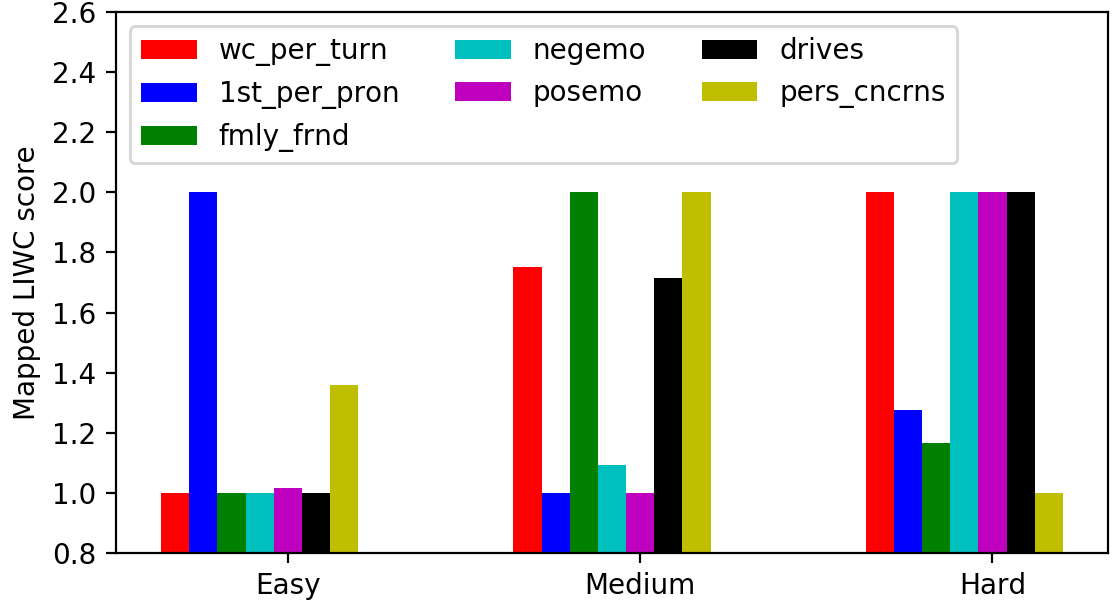}
  \caption{SD cues for three topic classes, mapped to interval [1,2] for better visualization}
  \label{fig:MappedLIWCscores}
\end{figure}

\begin{table}[t!]
\centering
\scalebox{0.9}{
\begin{tabular}{|l|l|}
\hline
  \textbf{Feature} & \textbf{Highest score sessions} \\
  \hline
  1st per. pron. & Getting to know,Yourself,Family \\
  \hline
  Fmly/frnd & Gathering, Family, Cooking   \\
  \hline
  Neg. emot. & Driving, Growing older, Money  \\
  \hline
  Pos. emot. & Yourself, Weather, Outdoors  \\
  \hline
  Drives & Gathering, Life goals, Arts  \\
  \hline
  Pers. concern  & Growing older, Activity, Family \\
  \hline
\end{tabular}}
\caption{\label{sd-cues-topics-table} Topics with the highest LIWC category scores}
\end{table}

Under this heading we focus on sessions mainly concerning user's lives, beliefs, interests, etc., expected to elicit some degree  of self-disclosure. The goal is to gain insight into the dependence of self-disclosure on different topics. As mentioned earlier, there is no well-defined set of cues for measuring self-disclosure, but various studies have suggested some potentially significant ones (recall section \ref{sec:background}). We instantiated these as follows, relying on LIWC features \cite{pennebakerdevelopment}: 
1) word count per turn, 2) first person pronoun, 3) family and friends, 4) negative emotions (anxiety, anger and sadness), 5) positive emotions, 6) drives (affiliation, achievement, power, reward, risk), 7) personal concerns (work, leisure, home, money, religion and death). 

We first report the LIWC-based scores of the above features in the three topic classes in table \ref{sd-cues-score-table}. To make the comparison more vivid, we linearly map the scores to [1,2] for each category independently and plot a bar graph. It can be seen that the ``hard" topics contain more words per turn, and more negative and positive emotions and drives. On the other hand, people use personal pronouns more often in easy topics such as when they introduce themselves or talk about their activities. Conversation about family, friends, and personal concerns, though somewhat intimate, need not involve high self disclosure. 

We also make a list of topics with the highest LIWC category scores. As can be observed in table \ref{sd-cues-topics-table}, participants used the most first-person pronouns in the initial greeting session and in talking about themselves and their families. Family and friend words not surprisingly were used in ``Family" and ``Gathering" sessions but also when the topic was on ``Cooking". ``Growing older" is among the topics where people use the most negative emotion and personal concern words.


\section{Conclusion}

\vspace*{-.06in}

We  presented  some  results  concerning  the  dialogue  behavior and  inferred  sentiment  of a  group  of  older  adults  interacting  with  a computer-based avatar on a wide range of topics. The naturalness of the interactions, generally attested by the users,\cite{razavi2019dialogue} indicates that our results are meaningful.
We observed that people tend to talk more when the topics are more intimate, such as life goals and the challenges of getting older, where they also use stronger emotion words---both positive and negative. Furthermore, the average response length increases as people progress along the series of interactions. 
These results support the use of dialogue agents with older adults in the context of difficult conversation topics. Our participants were more engaged with the agent when the conversation topics were more emotionally intense and intimate. Given the importance of effective communication during challenging conversations in later life—---driving cessation, healthcare, and end-of-life decision-making---our findings suggest that dialogue agents could provide valuable practice and coaching to help older adults successfully navigate these challenging conversations and thereby improve both health and quality of life.

Larger studies, and branching out to other age and culture groups, will be needed to gain a fuller understanding of user behavior in such settings, and to make inferences going beyond correlations to causal analyses.

\bibliography{acl2019}
\bibliographystyle{acl_natbib}

\end{document}